\documentclass[a4paper]{jpconf}
\usepackage{graphicx,cite}

\input{paperdef}

\begin{document}
\title{Electroweak Physics at the ILC}

\author{G.~Weiglein}

\address{
IPPP, Department of Physics, Durham University,
Durham DH1 3LE, U.K.
}

\ead{Georg.Weiglein@durham.ac.uk}

\begin{abstract}
Some aspects of electroweak physics at the International Linear Collider
(ILC) are reviewed. The importance of precision measurements in the
Higgs sector and in top-quark physics is emphasized, and the physics
potential of the GigaZ option of the ILC is discussed. It is shown in
particular that even in a scenario where the states of new physics are
so heavy that they would be outside of the reach of the LHC and the
first phase of the ILC, the GigaZ precision on the effective weak mixing
angle may nevertheless allow the
detection of quantum effects of new physics.
\end{abstract}


\section{Introduction}

The International Linear Collider (ILC) is a proposed electron--positron
collider
whose design is being addressed in the context of the Global Design
Effort~\cite{lcorg}. The ILC has been agreed in a world-wide consensus 
to be the next large experimental
facility in high-energy physics (see \citere{BriefingBook} and
references therein). The Reference Design Report for the ILC has been
issued earlier this year~\cite{lcorg}, and the Engineering Design Report
is currently in preparation.

The baseline design of the
ILC foresees a first phase of operation with a tunable energy of up to
about 500~GeV and polarised beams. Possible options include running at
the Z-boson pole with high luminosity (GigaZ) and running in the
photon--photon, electron--photon and electron--electron collider modes. The
physics case of the ILC with centre-of-mass energy of 400--500~GeV rests
on high-precision measurements of the properties of the top quark at the
top threshold, the unique capability of performing a comprehensive
programme of precision measurements in the Higgs sector, which will be
indispensable to reveal the nature of possible Higgs candidates, the
good prospects for observing the light states of various kinds of new
physics in direct searches, and the sensitivity to detect effects of new
physics at much higher scales by means of high-precision
measurements~\cite{ilc}. 

The baseline configuration furthermore foresees the possibility of an
upgrade of the ILC to an energy of about 1~TeV. The final choice of the
energy and further possible machine and detector upgrades will depend on
the results obtained at the LHC and the first phase of the ILC.   

The information on TeV scale physics obtainable at the
electron--positron collider ILC will be complementary to the one from
the proton--proton collider LHC~\cite{LHC,lhcilc}. While the discovery of
new particles often requires access to the highest possible energies,
disentangling the underlying structure calls for highest possible
precision of the measurements. Quantum corrections are influenced by the
whole structure of the model. Thus, the fingerprints of new physics
often only manifest themselves in tiny deviations. While in hadron
collisions it is technically feasible to reach the highest
centre-of-mass energies, in lepton collisions (in particular
electron-positron collisions) the highest precision of measurements can
be achieved. High-precision physics at the ILC is made possible in
particular
by the collision of point-like objects with exactly defined
initial conditions, by the tunable collision energy of the ILC, and by
the possibility of polarising the ILC beams. Indeed, the machine running
conditions can easily be tailored to the specific physics processes or
particles under investigation. The signal-to-background ratios at the
ILC are in general much better than at the LHC. In contrast to the LHC,
the full knowledge of the momenta of the interacting particles gives
rise to kinematic constraints, which allow reconstruction of the final
state in detail. The ILC will therefore provide very precise
measurements of the properties of all accessible particles. 
Direct discoveries at the ILC will be possible up to the kinematic limit
of the available energy.  Furthermore, the sensitivity to quantum
effects of new physics achievable at the ILC will in fact often exceed
that of the direct search reach for new particles at both the LHC and
the ILC.

The ILC can deliver precision data obtained from running at the top
threshold, from fermion and boson pair production at high energies, from
measurements in the Higgs sector and of possible other new physics. 
Furthermore, running the ILC in
the GigaZ mode yields extremely precise information on the effective
leptonic weak mixing angle at the Z-boson resonance, $\sweff$,
and the mass of the W~boson, $\MW$ (the latter from running
at the WW threshold). The GigaZ running can improve the accuracy in the
effective weak mixing angle by more than an order of magnitude. The
precision of the W mass would improve by at least a factor of two
compared to the expected accuracies at the Tevatron and the LHC.
Comparing these measurements with
the predictions of different models provides a very sensitive test of
the theory~\cite{gigaz}, 
in the same way as many alternatives to the Standard Model
(SM) have been
found to be in conflict with the electroweak precision data in the past.

In the following, some examples of electroweak physics at the ILC are
briefly discussed.


\section{Higgs physics at the ILC}

The high-precision information obtainable at the ILC will be crucial for
identifying the nature of new physics.
For instance, once one or more Higgs candidates
are detected, a comprehensive programme of precision 
measurements will be necessary to reveal the properties of the new 
state(s) and to determine the underlying physics.
The mass of the Higgs boson can be
determined at the ILC at the permille level or better, Higgs couplings
to fermions and gauge bosons can typically be measured at the percent
level, and it will be possible to unambiguously determine the quantum
numbers in the Higgs sector. Indeed, only the ILC may be able to discern
whether a Higgs candidate observed at the LHC is the Higgs boson of the 
SM or a Higgs-like
(possibly composite) scalar tied to a more complex mechanism of mass
generation. The verification of small deviations from the SM may be the
path to decipher the physics of electroweak symmetry breaking. The
experimental information from the ILC will be even more crucial if the
mechanism of electroweak symmetry breaking in nature is such that either
Higgs detection at the LHC may be difficult or the Higgs signal, while
visible, would be hard to interpret. 

A possible scenario giving rise to non-standard properties of the Higgs
sector is the presence of large extra dimensions, motivated for instance
by a ``fine-tuning'' and ``little hierarchy'' problem of supersymmetric
extensions of the SM.
A popular class of such models comprise those
in which some or all of the SM particles live on 3-branes in the extra
dimensions. Such models inevitably require the existence of a radion
(the quantum degree associated with fluctuations of the distance
between the 3-branes or the size of the extra dimension(s)).
The radion has the same quantum numbers as a Higgs boson. As a
consequence, there will in general be a mixing between the Higgs boson(s) and
the radion. Since the radion has couplings that
are very different from those of the SM Higgs boson, the physical
eigenstates will have  unusual properties corresponding to
a mixture of the Higgs and radion properties. In such a situation the 
ILC could observe both the Higgs and the
radion and measure their properties with sufficient accuracy to
experimentally establish the Higgs-radion mixing effects. 

If no clear Higgs signal has been established at the LHC, it will be
crucial to investigate with the possibilities of the ILC whether the
Higgs boson has not been missed at the LHC because of its non-standard
properties. This will be even more the case if the gauge sector does not
show indications of strong electroweak symmetry breaking dynamics. The
particular power of the ILC is its ability to look for $e^+e^-\to ZH$
 in the inclusive $e^+e^-\to ZX$ missing-mass distribution recoiling 
against the Z~boson. Even if the Higgs boson decays in a way that is 
experimentally
hard to detect or different Higgs signals overlap in a complicated way,
the recoil mass distribution will reveal the Higgs-boson mass spectrum
of the model. The total Higgs-strahlung cross section will be measurable
with an accuracy of about 2.5\% for a Higgs boson with a mass of about 120~GeV.
Should no fundamental Higgs boson be discovered, neither at the LHC nor
at the ILC, high-precision ILC measurements will be a direct probe of
the underlying dynamics responsible for particle masses. The LHC and the
ILC are sensitive to different gauge boson scattering channels and yield
complementary information~\cite{lhcilc}.

\section{Top and electroweak precision physics}

The ILC is uniquely suited for carrying out high-precision top-quark
physics. The mass of the top quark, $\mt$, is a fundamental parameter of
the electroweak theory. It is by far the heaviest of all quark masses and
it is also larger than the masses of all other known fundamental
particles.
The large value of $\mt$ gives rise to a large coupling between the top
quark and the Higgs boson and is furthermore important for flavour
physics. The top quark
could therefore provide a window to new physics. The correct
prediction of $\mt$ will be a crucial test for any fundamental theory.
The top-quark mass also plays an important role in electroweak precision
physics, as a consequence in particular of non-decoupling effects being
proportional to powers of $\mt$. A precise knowledge of $\mt$ is
therefore indispensable in order to have sensitivity to possible effects
of new physics in electroweak precision tests~\cite{deltamt}.

The ILC measurements at
the top threshold will reduce the experimental uncertainty on the
top-quark mass to the level of 100 MeV or below~\cite{ilc,mtdet}, 
i.e., more than an order
of magnitude better than at the Tevatron~\cite{mtTev} and the 
LHC~\cite{LHC,mtdetLHC}, 
and would allow a much more
accurate study of the electroweak and Higgs couplings of the top quark.
A precision of $\mt$ significantly better than 1~GeV will be necessary in
order to exploit the prospective precision of the electroweak precision
observables. In particular, an experimental error on $\mt$ of 0.1~GeV
induces an uncertainty in the theoretical prediction of $\MW$ and the
effective weak mixing angle, $\sweff$, of 1~MeV and $0.3 \times 10^{-5}$, 
respectively~\cite{deltamt}, i.e.,
below the anticipated experimental error of these observables. 

The impact of the experimental error on $\mt$ is even more pronounced in Higgs
physics.
In each model where the Higgs-boson mass is not a free
parameter but predicted in terms of the other model parameters
(as, e.g., in supersymmetry) the leading top-quark loop contribution
induces a correction to the Higgs-boson mass of the form
\BE
\De\mh^2 \sim \gf \; N_C \; C \; \mt^4~.
\EE
Here $\gf$ is the Fermi constant, $N_C$ is the colour factor, and the
coefficient $C$ depends on the specific model. Taking the Minimal
Supersymmetric Standard Model (MSSM) as an example
(including also the scalar top contributions and the appropriate
renormalisation) $N_C \, C$ is given for the light $\cp$-even Higgs
boson mass by
\BE
N_C \, C = \frac{3}{\sqrt{2}\,\pi^2\,\SQb} \;
\log \KL \frac{\mste\mstz}{\mt^2} \KR~.
\EE
Here $m_{\tilde t_{1,2}}$ denote the two masses of the scalar tops.
An LHC precision of $\de\mt = 1 \gev$ leads to an uncertainty of
the prediction for $\mh$ induced by $\de\mt$ of also about $1 \gev$.
The ILC accuracy on $\mt$, on the other hand, will yield a
precision of the Higgs-mass prediction of about $0.2 \gev$ 
(assuming that uncertainties from unknown
higher-order corrections can be brought sufficiently well under 
control~\cite{mhiggsAEC}).
These uncertainties have to be compared with the anticipated precision
of the measurement of the mass of a light SM-like Higgs boson at the
LHC, $\de\mh^{\rm exp,LHC} \approx 0.2 \gev$~\cite{LHC}. Thus, the ILC
precision on
$\mt$ is mandatory in order to obtain a theoretical prediction for $\mh$
with the same level of accuracy as the anticipated experimental
precision on the Higgs-boson mass.


\section{Electroweak precision observables in the MSSM}
\label{sec:zobs}

The high-precision measurement of the effective
leptonic weak mixing angle at the Z-boson resonance, $\sweff$, at GigaZ
provides an extremely sensitive probe of quantum effects of new
physics~\cite{gigaz}. In \citere{zobs} precision physics at the Z-boson
resonance has been discussed in the context of the MSSM, based on
state-of-the-art theoretical predictions. It has been analysed in
particular whether the high accuracy achievable at
the GigaZ option of the ILC would provide sensitivity to indirect effects of
SUSY particles even in a scenario where the (strongly interacting) 
superpartners are so heavy that they escape detection at the LHC.

\begin{figure}[bth!]
\begin{center}
\includegraphics[width=.6\textwidth]{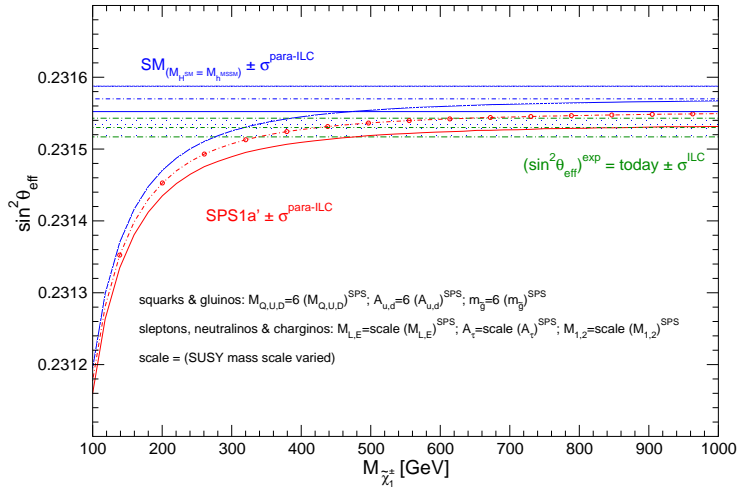}
\vspace{-1.0em}
\caption{
Theoretical prediction for $\sweff$ in the SM and the MSSM (including
prospective parametric theoretical uncertainties) compared to
the experimental precision at the ILC with GigaZ option.  
An SPS1a$'$ inspired scenario is used, where the squark and gluino
mass parameters
are fixed to 6~times their SPS~1a$'$ values. The other mass 
parameters are varied with a common scalefactor.}
\label{fig:ILC} 
\end{center}
\end{figure}

In \reffi{fig:ILC} a scenario with very heavy squarks and a very heavy
gluino is considered. It is based on the values of the SPS~1a$'$ benchmark
scenario~\cite{sps}, but the squark and gluino
mass parameters
are fixed to 6~times their SPS~1a$'$ values. The other masses are 
scaled with a common scale factor 
except $\MA$, the mass of the $\cp$-odd Higgs boson,
which is kept fixed at its SPS~1a$'$ value.
In this scenario 
the strongly interacting particles are too heavy to be detected at the
LHC, while, depending on the scale-factor, some colour-neutral particles
may be in the ILC reach. \reffi{fig:ILC} shows the prediction for
$\sweff$ in
this SPS~1a$'$ inspired scenario as a function of the lighter chargino
mass, $\mcha{1}$. The prediction includes the parametric
uncertainty, $\si^{\rm para-ILC}$, induced by the ILC measurement of $\mt$, 
$\de\mt = 100 \mev$, and the numerically more
relevant prospective future uncertainty on $\De\al^{(5)}_{\textup{had}}$,
$\de(\De\al^{(5)}_{\textup{had}})=5\times10^{-5}$~\cite{fredl}. 
The MSSM prediction for $\sweff$
is compared with the experimental resolution with GigaZ precision,
$\si^{\rm ILC} = 0.000013$, using for simplicity the current
experimental central value. The SM prediction (with 
$\MHSM = \Mh^{\rm MSSM}$) is also shown, applying again the parametric 
uncertainty $\si^{\rm para-ILC}$.

Despite the fact that no coloured SUSY 
particles would be observed at the LHC in this scenario, the ILC with
its high-precision 
measurement of $\sweff$ in the GigaZ mode could resolve indirect effects
of SUSY up to $m_{\tilde\chi^\pm_1} \lsim 500 \gev$. This means that the
high-precision measurements at the ILC with GigaZ option could be
sensitive to indirect effects of SUSY even in a scenario where SUSY
particles have {\em neither \/} been directly detected at the LHC nor the
first phase of the ILC with a centre of mass energy of up to $500 \gev$.



\subsection*{Acknowledgements}

The author thanks the authors of \citeres{lhcilc,BriefingBook,zobs}
for collaboration on the topics discussed in this paper and 
G.~Moortgat-Pick for interesting discussions concerning 
\refse{sec:zobs}. 
Work supported in part by the European Community's Marie-Curie Research
Training Network under contract MRTN-CT-2006-035505
`Tools and Precision Calculations for Physics Discoveries at Colliders'
(HEPTOOLS). 

\vspace{1em}



\end{document}